\documentclass[10pt,letterpaper]{article}
\usepackage{opex3}
\usepackage{cite}

\begin{document}
                       
\title{Cavity optoelectromechanical regenerative amplification}  
\author{Michael~A.~Taylor,$^1$ Alex~Szorkovszky,$^1$ Joachim~Knittel,$^1$ Kwan~H.~Lee,$^2$ Terry~G.~McRae,$^1$ and Warwick~P.~Bowen$^{1*}$} 
\address{$^1$ Centre for Engineered Quantum Systems, University of Queensland, St Lucia, Queensland 4072, Australia \\ $^2$ School of Mathematics and Physics, University of Queensland, St Lucia, Queensland 4072, Australia}
\email{$^*$ wbowen@physics.uq.edu.au}   
\date{\today}


\begin{abstract}
Cavity optoelectromechanical regenerative amplification is demonstrated. An optical cavity  enhances mechanical transduction, allowing sensitive measurement even for heavy oscillators. A 27.3~MHz mechanical mode of a microtoroid was linewidth narrowed to $6.6 \pm 1.4$~mHz, 30 times smaller than previously achieved with radiation pressure driving in such a system.  These results may have applications in areas such as ultrasensitive optomechanical mass spectroscopy.
\end{abstract}




\section{Introduction}

High quality factor ($Q$), low linewidth mechanical oscillators have many applications, such as highly sensitive spin~\cite{Rugar} or charge sensing~\cite{Cleland}, frequency standards in clocks~\cite{Major}, mass sensing with subzeptogram sensitivity~\cite{Jensen}, characterizing surface diffusion processes~\cite{Yang} and measuring forces with attonewton resolution~\cite{Teufel}.  The linewidth of a mechanical mode is determined by the energy dissipation rate of the mode, which without feedback is dominated by friction forces~\cite{Anetsberger2008}. A common technique to reduce the linewidth is to apply a feedback force to amplify the oscillator motion and bring it into the regenerative oscillation regime. When the feedback force overcomes friction, the oscillation becomes coherent and self-sustained. This reduces the mechanical linewidth and increases the mechanical energy by several orders of magnitude, which facilitates extremely precise monitoring of mechanical frequency-shifts.

The past few years have seen the rapid development  of cavity optomechanical systems, which combine mechanical oscillators with a confined optical field. In these systems, the strong interaction between the mechanical motion of the resonator and the cavity-enhanced optical field allows both control and measurement of the motion. Cavity optomechanical systems have many promising applications, such as on-chip phononic information processing~\cite{Lin,Naeini}, displacement sensing at the standard quantum limit~\cite{Anetsberger,Schliesser}, and ultrasensitive force measurement~\cite{Gavartin}.
 Ground-state cooled optomechanical oscillators are also proposed to probe exotic problems such as macroscopic quantum behavior~\cite{Arndt}, quantum gravity~\cite{Marshall} and microscale gravity~\cite{Haiberger}.

Radiation pressure driven regenerative amplification was demonstrated in an early cavity optomechanical experiment~\cite{Kippenberg}. However, the process is constrained by the low magnitude of available radiation pressure forces and the inability to modify the force spectrum independently from optical parameters. This both limits the achievable mechanical line narrowing, and results in unavoidable mechanical frequency pulling which introduces noise in sensing applications~\cite{Mirzaei}.  These constraints are overcome here using electrical gradient forces and ultrasensitive optomechanical transduction within a feedback loop, rather than radiation pressure, to control the mechanical motion.
%
In the context of cavity optomechanics, such a control system was first implemented in a recent demonstration of optoelectromechanical  feedback cooling of a microtoroid~\cite{Lee_K,McRae}, and has also been demonstrated in silicon disk resonators~\cite{Sridaran}.
 Some applications require large oscillators, such as mass sensing of large samples. However, transducing the motion of large oscillators can be difficult as mechanical amplitude scales inversely with the square root of resonator mass. In this regime, cavity enhanced optical readout is particularly relevant as it offers extremely sensitive detection. For applications requiring high sensitivity and large size, cavity optoelectromechanical oscillators are a promising class of resonators. 
Note that, similar to regenerative amplification, mass sensing can be achieved via mechanical parametric amplification~\cite{Zhang}. This is a fundamentally different process based on electrically induced mechanical nonlinearities, which by contrast is phase sensitive and requires no feedback~\cite{Rugar1991}.


 In this work, a model of the regenerative amplification process is derived which includes both electrical and radiation pressure forces. We experimentally demonstrate regenerative oscillations in a silica microtoroidal optoelectromechanical system, observing mechanical linewidths as low as $6.6 \pm 1.4$~mHz with a corresponding effective mechanical quality factor of $4\times10^9$. 
 For comparison, the best optically driven oscillators published have achieved 200~mHz linewidth in a microtoroid~\cite{Zadeh}, and recently 20~mHz in a silicon nitride ring oscillator~\cite{Tallur}.
  Recent experiments in a silicon oscillator have also demonstrated cavity optoelectromechanical regenerative amplification, as a means to filter and modulate the optical fields~\cite{Sridaran2011}. The higher optical quality factor of reflown silica used in the work reported here, combined with externally controllable electrodes allow 15~dB lower phase noise to be achieved at similar offset frequencies. This is an enabling step towards a new range of applications, including ultrasensitive mass spectroscopy~\cite{Feng,Naik}, photonic clocks~\cite{Zadeh} and a range of nonlinear radio-frequency (RF) processes~\cite{Zadeh2010}.

\begin{figure}
 \begin{center}
   \includegraphics[width=11cm]{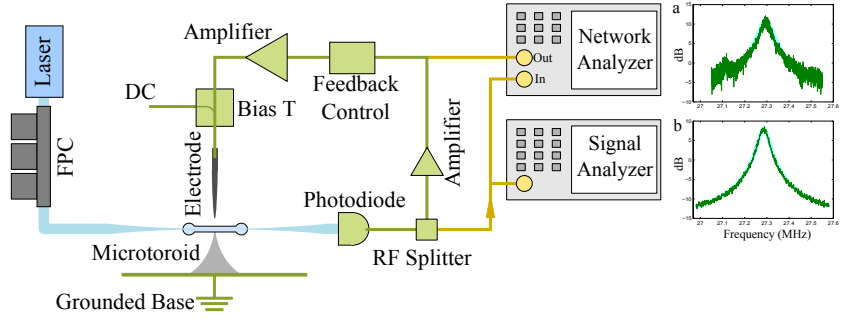}
   \caption{A schematic of the experiment.  FPC, fiber polarization control. The Feedback Control includes filtering and control over both feedback phase and amplitude. The network analyzer was used to characterize the driven response of the system, and all other results were taken with the signal analyzer. The subplots show the motion of the mechanical mode investigated here, (a) with driving from the network analyzer, and (b) the thermal motion measured on the spectrum analyzer. Shown in light blue is a fit to the data shown in green, which determines the  intrinsic decay rate $\Gamma_0$ and frequency $\omega_m$. The vertical axis of each subplot is in dB against an arbitrary reference.}
 \label{layout} 
 \end{center}

\end{figure}

\section{Theory}
 
The motion of a mechanical oscillator under the action of thermal $F_T$, optical $F_{\rm opt}$, and feedback $F_{fb}$ forces can be described by the equation of motion
\begin{equation}
m[\ddot {x}(t) + \Gamma_0 \dot {x}(t) + \omega_m^2 x(t)] =F_{fb}(t)+F_{\rm opt}+F_T(t) \label{Eq1},
\end{equation}
where $m$, $\Gamma_0$ and  $\omega_m$ are respectively the effective mass, dissipation rate and natural frequency~\cite{Carmon2005}. When the feedback force is proportional to the velocity, it opposes the dissipation, and if large enough, will cause regenerative oscillations. In the recent work of Kippenberg {\em et al.}~\cite{Kippenberg}, optical radiation pressure was used to drive an oscillator to regenerative amplification.
The radiation pressure force due to intracavity optical power $P_{IC}$ is given by $F_{\rm opt}= \frac{2\pi n}{c m} P_{IC}$, with $n$ the index of refraction~\cite{Rokhsari}. With the optical power below the threshold for regenerative amplification, this modifies the equation of motion to  
\begin{equation}
m[\ddot {x}(t) + \Gamma_r \dot{x}(t) + \omega_r^2 x(t)] =F_{fb}(t)+F_T(t) \label{RadPress},
\end{equation}
where the action of the optical force is to decrease the mechanical decay rate to $\Gamma_r=\Gamma_0 (1-  \frac{P_{\rm opt}}{P_{\rm thresh}})$ and shift the mechanical resonance frequency to $\omega_r=\omega_m (1+\eta_p P_{\rm opt}) $~\cite{Zadeh}. The frequency pulling is unavoidable with optical driving, and could limit sensitivity in some applications such as mass sensing. The regenerative oscillation regime is entered when the incident optical power $P_{\rm opt}$ exceeds the threshold  power $P_{\rm thresh}$, such that the decay rate becomes negative. The degree of frequency pulling is determined by the constant $\eta_p$, which depends on the optical detuning and physical parameters of the system.

In this article we include electrical feedback while keeping the optical power below the threshold for optical regenerative amplification. Assuming the mechanical oscillations shift the optical resonant frequency much less than the optical linewidth, the mechanical transduction is linear. Then the applied force is $F_{fb}(t)=m \omega_m \Gamma_r G x(t-\tau )$, where $\tau$ is a small delay such that $x(t-\tau) \propto \dot {x}(t)$, $G=G(x,G^0)$ is the steady-state feedback gain and $G^0$ is the small-signal gain.  
 The gain $G$ is normalized here so that regenerative amplification occurs for $G^0>1$.  Below this threshold there is no saturation in the feedback, and the gain is given by $G=G^0$.

  In the regenerative amplification regime, the feedback exceeds the dissipation, and the motion grows exponentially. The feedback force also grows exponentially as it is proportional to the position, until a component of the feedback electronics saturate, causing the steady-state gain $G$ to take a value which is very close to but smaller than 1. To determine the narrowed mechanical linewidth, we substitute the expression for the electrical feedback force into Eq.~(\ref{Eq1}), Fourier transform and rearrange to find
\begin{equation}
x(\omega) = \frac{F_T(\omega)}{m(\omega_r^2 - \omega^2 +  i \Gamma_r \omega) -i m \omega_m \Gamma_r G } \label{Eq2}.
\end{equation}
 In the high $Q$ limit, the frequency range of interest lies very near the mechanical resonance frequency. We therefore take $\omega=  \omega_r+\Delta$, with $\Delta \ll \omega_m$, and assume perfect feedback phase such that $e^{-i\omega \tau} \approx i+\Delta \tau$. With these approximations, Eq.~(\ref{Eq2}) may be re-expressed as 
\begin{equation}
x(\Delta) =\frac{1}{m \omega_m} \, \frac{F_T(\omega)}{-2  \Delta  + i \Gamma_r (1-G) }. \label{Spectrum}
\end{equation}
The feedback narrowed full-width half-maximum linewidth $\Gamma$ is then easily found to be
\begin{equation}
\Gamma = \Gamma_0 (1-  \frac{P_{\rm opt}}{P_{\rm thresh}}) (1-G) \label{LW_g}.
\end{equation}
 We see that the linewidth $\Gamma$ goes to zero as the gain $G$ approaches 1 or the optical power $P_{\rm opt}$ approaches the threshold $P_{\rm thresh}$. The steady state gain $G$ can be found in terms of experimentally measurable parameters by calculating the total mechanical energy $E_{\rm osc}$
\begin{equation}
E_{\rm osc} =  \frac{1}{2\pi} m \omega_m^2 \int_{-\infty} ^{\infty} \left| x(\omega)\right| ^2 d\omega 
 =\frac{\frac{1}{2}k T}{(1-  \frac{P_{\rm opt}}{P_{\rm thresh}})(1-G)} \label{E_g},
\end{equation}
where $\left| F_T (\omega)\right| ^2  = m \Gamma_0 k T$ due to the fluctuation-dissipation theorem, with thermal amplitude fluctuations freezing out in the high power limit~\cite{Rokhsari,Edson}. In the absence of driving ($G=0$ and $P_{\rm opt}=0$) these fluctuations do not freeze out, doubling the thermal contribution and reducing this expression to the expected thermal energy $E_T = kT$.
Eq.~(\ref{E_g}) allows the steady-state gain $G$ to be established from measurements of the intrinsic mechanical decay rate and the oscillator energy with and without feedback. $G$ approaches unity in the limit of regenerative amplification, with $1-G \approx 10^{-5}$ in the experiments reported here. The oscillator energy can then be expressed simply as $\frac{E_T}{E_{\rm osc}}=2(1-G)(1-  \frac{P_{\rm opt}}{P_{\rm thresh}})$. Substituting this expression into Eq.~(\ref{LW_g}) finally gives the expected narrowed mechanical linewidth.
\begin{equation}
\Gamma= \frac{\Gamma_0}{2} \frac{E_T}{E_{\rm osc}}.\label{LWeq1}
\end{equation}
For optical powers below the threshold for radiation pressure driven regenerative oscillations, as in our experiment, optical driving is indistinguishable from electrical feedback in this expression. This expression for the linewidth is identical to that obtained from a more complex analysis of radiation pressure driving~\cite{Rokhsari}, and is of a similar form to those for other regenerative amplifiers, such as RLC electronic oscillators~\cite{Robins}, optoelectronic oscillators~\cite{Yao}, masers~\cite{Gordon}, and the Schawlow-Townes limit for  laser linewidth~\cite{Schawlow,Paschotta}.
Note that an equivalent expression could be derived in terms of the mechanical output power rather than the oscillator energy, similar to the expressions common for lasers and optoelectronic oscillators. Here, however, the oscillator energy is more generally relevant, since in contrast to those situations, the output phonon field is not readily accessible. It is instead enhancements in the oscillator energy upon which applications typically depend.


\section{Experiment and results}

\subsection{Experimental system}

To experimentally demonstrate optoelectromechanical regenerative amplification, a room temperature silica microtoroid was used with electrical gradient forces provided by an electrode placed in close proximity. The optical cavity arises due to light being confined in whispering-gallery modes within the toroid by total internal reflection. The mechanical oscillator is simply the natural vibrational modes of the physical structure. These vibrational modes modulate the path length of the optical cavity, which shifts the optical resonant frequency giving strong coupling between oscillator position and optical fields.  Silica microtoroids can have optical $Q$ factors of $10^8$, which allows motion transduction sensitivity at the level of $10^{-19}$~m~Hz$^{-1/2}$~\cite{Anetsberger}.

\begin{figure}
 \begin{center}
   \includegraphics[width=7cm]{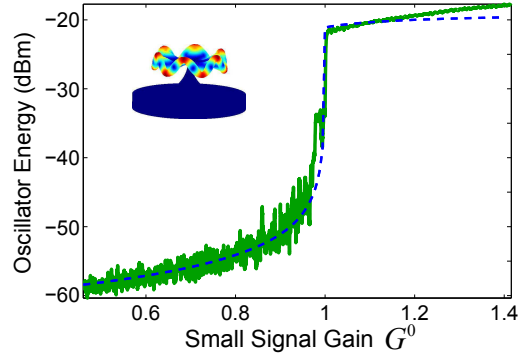}
   \caption{Oscillator mechanical energy as a function of small-signal feedback gain. Green trace: measured data. Below saturation the theoretical model (blue dashed line) is given by Eq.~(\ref{E_g}). To estimate the above-threshold mechanical energy, Eq.~(\ref{Eq1}) is numerically solved for sinusoidal motion with a limit applied to the magnitude of the feedback force $F_{fb}(t)$. Two fitting parameters are used; one normalizes the gain such that saturation occurs at $G^0=1$, and another defines the mechanical energy at which saturation occurs. Inset: A finite-element model of the mechanical mode being amplified.}
 \label{SmallSig} 
 \end{center}
 \end{figure}

A schematic of our experimental setup is shown in Fig.~\ref{layout}. A shot-noise limited fiber laser provided 60~$\mu$W of 1560~nm light, which coupled evanescently from a tapered optical fiber into a whispering-gallery optical mode with an intrinsic $Q$ factor of $3\times 10^6$, and with major and minor diameters of $65\mu$m and $6\mu$m.
The laser was blue-detuned to measure the motion of the microtoroid, so that mechanically induced optical resonant frequency shifts cause intensity modulation of the transmitted optical field. Thus mechanical motion was measured directly on the output light intensity with transduction sensitivity of $3\times10^{-17}$~m~Hz$^{-1/2}$. 
 We chose to study a 6th order crown mode at 27.3~MHz (shown in Fig.~\ref{SmallSig} inset), as it was clearly separated in frequency from other detected modes. Due to the symmetry of the structure, two spatially separated modes exist at this frequency. The mode which will be regeneratively amplified is defined spatially as the one which responds most to the electric force profile, while its orthogonal counterpart is not amplified. The response of this mode to coherent electrical driving from a network analyzer is shown in Fig.~\ref{layout}(a), demonstrating that this mode responds to the electrical driving. The thermal motion of this mode is shown in Fig.~\ref{layout}(b) to which we fit the mechanical parameters and find the mechanical quality factor to be $Q=600$.

 To enable sub-Hz linewidth measurements it was critical to suppress low frequency noise in the apparatus.  The primary sources of low frequency noise were motion of the tapered fiber due to air currents and thermal fluctuations in the toroid due to absorption of background light.  Air currents were eliminated by placing the experiment in a hermetically sealed box, and background light levels were minimized. The use of a shot noise limited laser and low noise electronics ensured that both the radiation pressure back-action on the mechanical oscillator and the electronic feedback were at the quantum noise limit. The low noise performance allows high sensitivity measurements with low optical intensity, thus minimizing optical back-action.
 The detected photocurrent was band-pass filtered to suppress other mechanical modes.
Electronic components controlled the phase (JSPHS-26) and amplitude (ZX73-2500) of this signal before it was amplified and fed back to an electrode. This amplified signal was fed back to a sharp stainless steel electrode with a 2 $\mu$m diameter tip.
 This allowed electrical gradient forces to be applied to the microtoroid following the approach of Ref.~\cite{Lee_K,McRae}.
  This force was intensified by placing the electrode at the end of a coaxial half-wave transmission line resonator tuned to the mechanical resonance frequency. This enhanced the applied force by the resonator's $Q$ factor of 25 compared to the work of Ref.~\cite{Lee_K,McRae}. Further to this, a constant voltage of 120~V was applied to the electrode increasing the polarization of the silica microtoroid structure and thus enhancing its response to the applied electric field. This constant polarization had a negligible effect on the mechanical mode properties.

The electrode was positioned 5--20 $\mu$m vertically above the microtoroid rim. At this location, it produced an electric field gradient which was not azimuthally symmetric, and therefore able to excite the mechanical mode. While the overlap between the applied force and the mechanical mode was small, it was sufficient to allow regenerative amplification of the mechanical mode. If the electric field were tailored to the mechanical mode, the same applied forces could be generated with much a lower electrode voltage.  This could be important in applications for which power consumption is a limiting factor.

\begin{figure}
 \begin{center}
   \includegraphics[width=7cm]{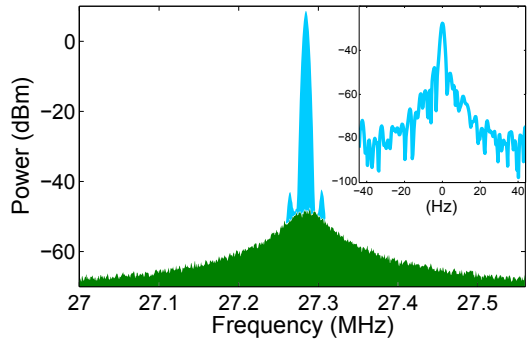}
   \caption{Blue and green traces respectively show the mechanical power spectrum with and without amplification. These were measured by sending the signal from the optical detector directly into a spectrum analyzer. Note that this mode is composed of two degenerate modes which are distinguished only by the azimuthal position of the nodes and antinodes. Only the mode with greatest overlap with the driving field is amplified, although both are present in the measurements.  Inset: A near resonance spectrum with amplification and a 1~Hz resolution bandwidth (right).}
 \label{Traces} 
 \end{center}
 
\end{figure}

 By adjusting the phase of the feedback, the gradient force was optimized to maximally amplify the mechanical motion of the 27.3~MHz mode, and the feedback gain was varied by adjusting the variable attenuator. As the small signal gain was increased toward threshold, the energy in the oscillator increased as expected from Eq.~(\ref{E_g}), and the mechanical resonance narrowed as described by Eq.~(\ref{LW_g}). Above threshold the feedback overcame the mechanical damping, with the mechanical energy growing until the feedback amplifier saturated. This clamped the energy to a fairly constant level, typically 4--5 orders of magnitude greater than the thermal energy. Fig.~\ref{SmallSig} shows the increase in oscillator energy as the small signal gain was increased across the threshold. 
 Spectra of the mechanical resonance are shown in Fig.~\ref{Traces} both without amplification, and in the regime of regenerative amplification. The amplified signal has a peak power spectral density which is over 50~dB higher and a sub-Hz linewidth, which could not be resolved on our spectrum analyzer (shown in inset).

\subsection{Determining the mechanical linewidth}

\begin{figure}
 \begin{center}
  \includegraphics[width=12cm]{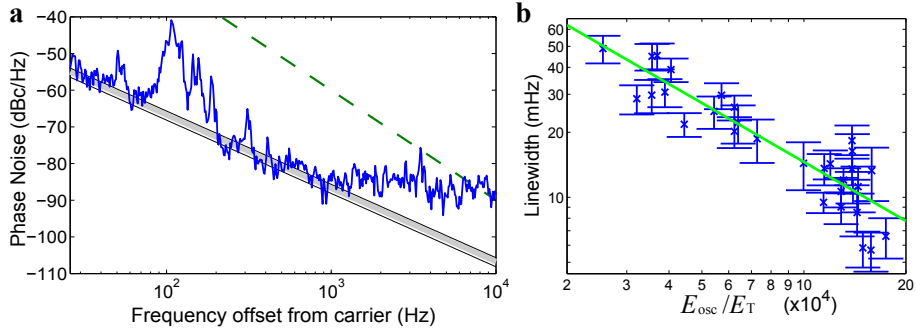}
   \caption{ (a) An example of a phase noise trace. The fitted noise floor is for a linewidth $\Gamma =13.3 \pm 3.7$~mHz. A low $Q$ mechanical resonance at 108~Hz is clearly visible, and electronic noise from the signal analyzer also appears for frequency offsets above 1~kHz. The dotted line indicates the phase noise typically achievable with optomechanical driving only~\cite{Zadeh}. (b) The measured linewidth as a function of oscillator energy. Fitting to the data (green) shows shows a dependence of $\Gamma \propto E_{\rm osc}^{-0.91\pm0.12}$, while theory predicts $\Gamma \propto E_{\rm osc}^{-1}$. }
 \label{LineWidth} 
 \end{center} 	
 
\end{figure}

Because the mechanical linewidth was unresolvable, it was determined via phase noise analysis. The finite linewidth of the mechanical signal causes a floor in phase noise power which is given by 
\begin{equation}
10^{\mathcal{L}(\Delta \Omega)/10}=\Gamma~ \Delta \Omega^{-2}, \label{PNeqn}
\end{equation}
 where $\Delta \Omega$ is the detuning from the carrier frequency and $\mathcal{L}(\Delta \Omega)$ is the phase noise power measured at the detuning normalized to the power of the central peak, in units of dBc/Hz~\cite{Paschotta,Rokhsari}. An example of a measured phase noise trace is shown in Fig.~\ref{LineWidth}(a). The results match the predicted $\Delta \Omega^{-2}$ dependence over a wide range of offset frequencies. At several frequencies the phase noise exceeds the prediction from the expression above, with other noise sources dominant.   Much of the additional phase noise seen in the trace at low offset frequencies are contributed by harmonics of one low $Q$ mechanical resonance at 108~Hz in the tapered optical fiber. A high noise floor is present partly because we use a signal analyzer to extract phase noise rather than a dedicated phase noise analyzer. This noise floor was found to be well below the measured noise between 20~Hz and 1~kHz offset, where we perform our analysis. The minimum phase noise achieved in our system was determined to be $-84.4 \pm 1.1$ dBc/Hz at an offset frequency of 500~Hz.


The mechanical linewidth was inferred from phase noise data by fitting a noise floor proportional to $\Delta \Omega^{-2}$ to the phase noise spectrum, as in Fig.~\ref{LineWidth}(a). Since the measured phase noise must always be greater than or equal to the noise floor of Eq.~(\ref{PNeqn}), this fit gives an upper bound on linewidth. Due to the clear $\Delta \Omega^{-2}$ trend observed over a significant frequency range in all phase noise traces analyzed, we are confident this corresponds closely to the actual linewidth. To verify this we experimentally confirmed the predicted dependence of linewidth on mechanical energy.

To test this dependence, the phase noise was analyzed for data with a range of mechanical energies. Mechanical energy was controlled by varying the electrode-microtoroid gap while keeping the feedback above threshold, which adjusts the force applied to the mechanical mode. The energy was measured on a spectrum analyzer for each data point, and normalized against measurements of the thermal motion to give the ratio $E_{\rm osc}/E_T$. Each energy measurement was made with a bandwidth larger than the linewidth, so that the spectral peak height would contain essentially all of the mechanical energy. The measured thermal motion was from two degenerate modes, so had an energy of $2 kT$. The extracted linewidths are shown against oscillator energy in Fig.~\ref{LineWidth}(b). The linewidth is found to follow a trend of $\Gamma \propto E_{\rm osc}^{-0.91\pm0.12}$, which to within error matches the predicted relation from Eq.~(\ref{LWeq1}) of $\Gamma \propto E_{\rm osc}^{-1}$.  The linewidth achieved for maximum oscillator energy was $6.6 \pm 1.4$~mHz, giving an effective $Q$ factor of $4 \times 10^9$, compared to the smallest published linewidth achieved by optomechanical driving in a microtoroid of 200~mHz~\cite{Zadeh}, with an equivalent $Q$ factor of $2.5 \times 10^8$.

\subsection{Application to mass sensing}

This system has potential application for mass sensing, with the regeneratively narrowed mechanical linewidth providing high sensitivity to mechanical resonance frequency shifts due to the deposition of small masses on the oscillator. The mass sensitivity is then limited by the accuracy with which the mechanical resonance frequency can be determined within the detection bandwidth $\Delta f$. In our current experiments, the limiting factor is drift in the position of both the optical taper, and hence coupling condition, and electrode cause feedback phase fluctuations which in turn result in fluctuations of the mechanical resonance frequency. However, it is possible to stabilize the position of both taper~\cite{Junge} and electrode using active and/or passive techniques. Assuming that the drift in mechanical resonance frequency over the oscillation lifetime can be made smaller than the narrowed linewidth, thermomechanical fluctuations place a fundamental sensitivity limit to mass sensing \cite{Ekinci}
\begin{equation}
\delta m_{\rm min} = 2 m_{\rm mode} \left(\frac{E_T}{E_{\rm osc}} \right)^{1/2} \left(\frac{\Gamma}{2\pi Q \omega_m} \right)^{1/2},
\end{equation}
for a mode mass $m_{\rm mode}$ and detection bandwidth $\Delta f = \Gamma/2\pi$. The mode mass here is the mass which moves in the oscillations, $m_{\rm mode} = \int \rho |u({\bf r})|^2 d{\bf r}$, with $\rho$ the density and $u({\bf r})$ the spatial shape of the mode~\cite{Pinard}. Using the spatial profile calculated with finite-element modeling this gives a mode mass of $5\times10^{-9}$~g. Using this, along with $\frac{E_{\rm osc}}{E_T}=10^5$, $\Gamma= 2\pi\times0.01$~Hz, $Q=600$ and $\omega_m=2\pi\times27.3$~MHz, an optimum sensitivity of $\delta m_{\rm min} \approx 10^{-17}$~g is predicted at room temperature and atmospheric pressure.
%
%
%
%
%
%
 The predicted mass sensitivity is comparable to the the best sensitivities achieved in room temperature mechanical sensors. Sensitivity at this level has been achieved in microfluidic channels embedded within cantilevers~\cite{Lee_J}, and cantilevers in air~\cite{Verd}. Both of these systems have significant differences to our setup. The microfluidic integrated cantilevers measure sample masses within fluid, whereas the microtoroid would operate on samples in air. Cantilever operating in air perform a similar measurement, but are much smaller; the device in Ref.~\cite{Verd} is 3 orders of magnitude lighter than our microtoroid.  Performing mass measurement with the larger structure of a microtoroid may enable measurements with comparative sensitivity on substantially larger samples. This could allow, for example, sensitive characterization of single-cell dynamical mass changes in processes such as photosynthetic growth. Introducing cavity enhanced transduction to mechanical mass sensors could be a promising technique to extend the range of ultrasensitive measurements to larger samples.

\section{Conclusion}

Regenerative oscillation was achieved in a cavity optoelectromechanical system using feedback which combined cavity enhanced optical transduction with electrical actuation. The optical cavity enhances the mechanical transduction over standard optical or electrical techniques, and the electrical forces allow stronger force with improved control when compared to optomechanical driving of such an optomechanical cavity. A theoretical model of this system was formulated to include both radiation-pressure and electrical feedback. Linewidth narrowing from 46~kHz to $6.6 \pm 1.4$~mHz was achieved at a frequency of 27.3~MHz, corresponding to an effective quality factor of $4 \times 10^9$. This linewidth is smaller than that achieved in similar radiation pressure driven regenerative oscillators, which have been reported to reach 200~mHz. The linewidth was predicted and experimentally confirmed to scale inversely with mechanical energy. This opens new possibilities for sensitive mass spectroscopy at room pressure and temperature.

\section*{Acknowledgments}

This research was funded by the Australian Research Council Centre of Excellence CE110001013 and Discovery Project DP0987146.  Device fabrication was undertaken within the Queensland Node of the Australian National Fabrication Facility (ANFF).

\end{document}